\documentstyle{article}
\begin{document}
\noindent
{\bf QUASI-SET THEORY FOR BOSONS AND FERMIONS:} 

\vspace{1mm}
\noindent
{\bf QUANTUM DISTRIBUTIONS}

\renewcommand{\thefootnote}{\fnsymbol{footnote}}
\vspace{2.5cm}
\hspace{2.5cm}

\hspace{25mm} {\bf D. Krause$^*$,} \hspace{1mm} {\bf A. S. Sant'Anna}$^*$ 
\hspace{1mm} 
{\bf and}

\hspace{25mm} {\bf A. G. Volkov}\footnote{Dep. Matem\'atica, UFPR, C.P. 19081, Curitiba, PR, 81.531-990, Brazil. All correspondence should be sent to Dr. Adonai S. Sant'Anna. Phone: +55-41-246-3518; FAX: +55-41-267-4236; e-mail: adonai@mat.ufpr.br}

\vspace{0.5cm}
\hspace{2.5cm}

\vspace{1.0cm}

\hspace{2.5cm}\parbox{10.5cm}{
{\underline{\hspace{10.5cm}}\\
Quasi-set theory provides a mathematical background for dealing with collections of indistinguishable elementary particles. In this paper, we show how to obtain the quantum statistics into the scope of quasi-set theory and discuss the Helium atom, which represents the simplest example where indistinguishability plays an important role. A brief discussion about indistinguishability and interference is also presented as well as other related lines of work. One of the advantages of our approach is that one of the most basic principles of quantum theory, namely, the Indistinguishability Postulate, does not need to be assumed even implicitely in the axiomatic basis of quantum mechanics.}
\underline{\hspace{10.5cm}}}

\newtheorem{definicao}{\bf Definition}
\newtheorem{teorema}{\bf Theorem}
\setlength{\unitlength}{1mm}
\renewcommand{\thefootnote}{\arabic{footnote}}
\setcounter{footnote}{0}
\newcounter{cms}
\setlength{\unitlength}{1mm}

\pagestyle{myheadings}
\markright{Quasi-set theory for bosons and fermions}

\paragraph{1. INTRODUCTION}

\vspace{0.5cm}

\paragraph{ } The present paper has a philosophical purpose in the sense 
that it copes with philosophical questions regarding the foundations of 
quantum physics. On the other hand, it is also a work on mathematical 
physics, in the sense that we are concerned with the {\em mathematical} 
foundations of quantum physics. Mathematical physics should not be understood as regarding the study of mathematical techniques in theoretical physics only, but also as a discipline engaged in the mathematical foundations of physical theories. It is in this `foundational' sense that we are here studying some problems about the concepts of identity and individuality of the basic entities of quantum theory.

	It is well known that both classical and quantum particles which share the same set of intrinsic, state-independent properties like charge, rest-mass, etc., are {\it indistinguishable\/}.\footnote{We use the philosophical jargon in saying that `indistinguishable' objects are objects that share their properties, while `identical' objects are `the very same object'. See (Krause and French, 1995).} Nevertheless, there is a fundamental difference in the behaviour of the ensembles of such particles, as characterized by classical and quantum statistics respectively. In classical mechanics, Maxwell-Boltzmann statistics counts as distinct from the original one any arrangement obtained from a permutation of particles between states, contrarily to Bose-Einstein and Fermi-Dirac statistics, as it is well known. Then, it has been argued that classical particles are `individuals' of some kind and, since they are indistinguishable in the sense mentioned above, their individuality must be ascribed by something which `transcends' their properties; this idea has been discussed in (Krause and French, 1995; Post, 1963; Redhead and Teller, 1991; Redhead and Teller, 1992; Teller, 1995). In quantum statistics, on the other hand, the Indistinguishability Postulate asserts that ``If a permutation is applied to any state for an assembly of particles, then there is no way of distinguishing the resulting permuted state function from the original one by means of any observation at any time'' (French, Krause and Maidens, 1997). The Indistinguishability Postulate (henceforth `IP') is one of the most basic principles of quantum theory and implies that permutations of quantum particles are not regarded as observable (Greenberg and Messiah, 1964).

	Usually, IP has been interpreted according to two basic ways: the first one assumes that IP implies that quantum particles cannot be regarded as `individuals', since an `individual' should be something having properties similar to those of usual (macroscopic) bodies.\footnote{The concept of a `physical object' is difficult and controversial (Auyang, 1995; Castellani, 1997; Toraldo di Francia, 1978). In the papers by Redhead and Teller mentioned above there are additional discussions on this point.} This interpretation is closely related to what is assumed in the context of quantum field theory, since, roughly speaking, quantum field theories do not deal with `individuals' (cf. the above references). The second way in considering quantum particles regards them as individuals in a sense, and the non-classical counting of quantum statistics are then understood as resulting from the restrictions imposed to the set of the possible states acessible to the particles (French, 1989; French and Redhead, 1988; Redhead and Teller, 1992). In short, only symmetrical and antisymmetical states are available, and the initially attached individuality of particles is then `veiled' by such a criterion.

	Both alternatives mentioned above, albeit used in current literature, present problems from the `foundational' point of view. There is some obscurity lurking in the concept of individuality in quantum physics. The idea of considering `non-individuals' seems strong,\footnote{But in
(Krause and French, 1995) it was outlined a `rationale' for such a view.} and in general one prefers to use another metaphysical package, namely, that the quantum objects are individuals of a sort, despite quite distinct from the usual objects described by classical mechanics. The just mentioned papers provide a wider discussion on this topic, so we shall not recall all details here. We only remark that in the foundational studies on quantum theory, it is usual to consider questions of interpretation of the formalism instead of analysing the possibility of exchanging the underlying logico-mathematical apparatus (as observed in (Faris, 1996)). We prefer to follow this second approach. Thus, the present paper is the first one of a series devoted to such an exchanging of the mathematical framework of quantum theory. Although our attention in this paper is devoted mainly to quantum statistics, we recognize that there are other quantum phenomena where indistinguishability plays a fundamental role. But such questions will be discussed only in future papers.

	In a previous work, two of us have presented a manner to cope with collections of `physically' indistinguishable particles in a set-theoretical framework (that is, we have worked in standard set theory) by using hidden variables (Sant'Anna and Krause, 1997). Here, we use an alternative package, namely, quasi-set theory, to provide an adequate way to express the quantum 
statistics; it results that there is no necessity of the assumptiom that only `symmetrical' and `antisymmetrical' states are available (as implied by IP). This fact results as a natural consequence of our mathematical framework.

	In the next section, we recall the main features of a quasi-set theory. For the moment, let us say that by using quasi-set theory, we are not commited to the assumption that quantum particles are, at least in principle, individuatable\footnote{This term was coined by Redhead and Teller [op. cit.] to mean those entites which can be individualized by some device.} objetcs whose indistinguishability is ascribed only a posteriori by choosing the symmetric (and antisymmetric) vectors or, alternatively, the symmetric (and antisymmetric) solutions of the Schr\"odinger equation. Let us remark that by using classical mathematics (built on, say, Zermelo-Fraenkel set theory) to found quantum theory, we are necessarily commited in considering the basic entities as provided of individuality of some kind. In short, this is due to the fact that every entity is, at the end of the road, a set, and a set is, according to Cantor (1955), ``a collection into a whole of {\it distinct\/} objects of our intuition or of our thought'' (our emphasis).

	Some authors like Weyl expressed the calculation with `aggregates' so that the assumptions of quantum theory could be reached in a satisfactory way; but the only `objective' result of Weyl's efforts was to find an alternative manner to express the very same procedure physicists implicitely use, namely, the assumption that we have a {\it set\/} $S$ of (hence, distinguishable) objects (say, $n$ objects) endowed with an equivalence relation $\sim$. Then the `desired result', according to Weyl (1949), is to obtain the {\it ordered decomposition\/} $n = n_{1} + \ldots + n_{k}$, where $n_{i}$ are the cardinalities of the equivalence classes $C_{i}$, $i = 1, \ldots, k$ of the quotient set $S/\sim$. But, as it is easy to note, this procedure `veils' the very nature of the elements of the set $S$, that is, veils the fact that they are individuatable objects since they are members of a {\it set\/}.\footnote{One of us have analysed Weyl's ideas in connection with quasi-set theory; see (Krause, 1991).} We would like to emphasize that there is no scape. Classical logic and mathematics are commited with a conception of identity which does not make any distinction between identity and indistinguishability: indistinguishable things are the very same thing and conversely.

	By using quasi-set theory instead of standard set theory, our paper provides a way of obtaining the statistics in the direction traced by Post's (1963) suggestion that the `non-individuality' of quantum objects should be ascribed ``right at the start''; that is, contrarily to Weyl and the standard presentation of quantum theory, we keep them as objects devoided of identity just from the beginning. By this way, we invert the usual order of considering quantum entities; the indistinguishability among certain quantum objects is assumed as a primitive concept. The tools for considering that indistinguishability does not imply identity is provided by 
quasi-set theory. We also discuss the problem of the Helium atom, where indistinguishability should be assumed in order to define the wave function of such an atom.

\paragraph{1.1 SOME RELATED TOPICS}

\vspace{0.5cm}

\paragraph{ } There are of course several questions to be answered in 
connection with the discussion presented here. Let us give an idea of 
some of them, which despite their importance, will not be pursued in 
this paper.

	The first problem concerns the nature of the `properties' to be considered as {\it licit properties\/} of quantum objects. The question is subtle, and it is related with the concept of indistinguishability as presented above, for if indistinguishable particles are to be considered as those particles that share the same set of properties (of some kind), it seems clear that a reasonable definition of what is to be understood by a property should be provided. Some authors have mentioned the necessity of restricting the collection of properties to certain particular cases. Nevertheless, whatever definition we consider, it seems that there are only {\em ad hoc\/} reasons in the tentatives to dismiss some possible attribute of a thing as a legitime property of that thing, and the same happens in particular if the `thing' is an elementary particle. Let us be more precise on this point. When we consider Leibniz's Principle of the Identity of Indiscernibles (henceforth `PII'), which in a second order language with identity may be formulated as $$\forall P (P(x) \leftrightarrow P(y)) \to x = y,$$

\noindent
where $P$ is a variable that ranges over  the set of the attributes of $x$ and $y$, we must understand the quantifier $\forall$ as classical logic does (since we are supposing the valitity of classical logic by hyphotesis). That is, `forall' means for all, and not for some. This seems quite trivial, but without such a remark we cannot justify the restrictions of the range $P$ to
monadic properties only, or to relational properties only, or to spatio-temporal properties, as in (French and Redhead, 1988). In all these cases, it has not been assumed PII in full, since the range of $P$ was restricted. The possibility of admiting that there are various forms of PII sustains our belief that there is no reason to suppose that some kind of attribute, say relational properties, are not `licit'.

	Roughly speaking, {\it every\/} formula of an adequate language with just one free variable should stand for a `property'. By this way, even the `problematic' property $I_{a}$ of an object $a$ defined by $I_{a}(x) \leftrightarrow x = a$ must count as licit. The problem regarding $I_{a}$ is that if it is included in the class of $a$'s attributes, then every object
$b$ which share with $a$ all its properties (that is, every object indistinguishable from $a$) is so that $I_{a}(b)$ and hence it is identical with $a$. By this reason, it has been argued that $I_{a}$ is not a legitime property of $a$. Here, we do not need to make {\em ad hoc} restrictions on the set of possible properties of an elementary particle, since in quasi-set theory the lack of identity of some objects makes sense the fact that even a `property' like $I_a(x)$ should be considered as licit from the syntactical point of view. Nevertheless, it cannot be adequately attributed to $a$ since the concept of identity (represented here by the sign of equality) does not make sense for objects of the $a$-type. Important to note that this `restriction' results from a very natural way of considering particles, instead of a prejudice on the properties, and follows Sch\"odinger's ideas (da Costa and Krause, 1994). More details on this point can be found in (Krause, 199*b).

	Another point is that in considering quantum objects as devoided of well-defined identity, as quasi-set theory does, there remains the problem of explicating how a macroscopic body, which is composed by these entities, acquires its identity. Schr\"odinger (1952) expressed that in terms of a {\it Gestalt\/}, but of course we have a great problem in our hands. An interesting analysis of Schr\"odinger's ideas is presented in (Bitbol and Darrigol, 1992). We envisage that perhaps what is in need is a kind of `quantum mereology', that is, a logic of part-whole (Simons, 1987) suitable for quantum physics. Our suppositions here do not depend on this topic, which is also related with interesting points as, for instance, Gibb's paradox (Lesk, 1980). These topics will be pursued in future works.

\paragraph{2. OUTLINES OF THE THEORY ${\cal Q}$}

\vspace{0.5cm}

\paragraph{ } The quasi-set theory ${\cal Q}$ is based on Zermelo-Fraenkel-like axioms and allows the presence of two sorts of atoms ({\it Urelemente\/}), termed $m$-atoms and $M$-atoms.\footnote{All the details of this section may be found in (Krause, 199*a).} Concerning the $m$-atoms, a weaker `relation of indistinguishability' (denoted by the symbol $\equiv$), is used instead of identity, and it is postulated that $\equiv$ has the properties of an equivalence relation. The predicate of equality cannot be applied to the $m$-atoms, since no expression of the form $x = y$ is a formula if $x$ or $y$ denote $m$-atoms. Hence, there is a precise sense in saying that $m$-atoms can be indistinguishable without being identical. This justifies what we said above about the `lack of identity' to some objects.

	The universe of ${\cal Q}$ is composed by $m$-atoms, $M$-atoms and {\it quasi-sets\/} (qsets, for short). The axiomatics is adapted from that of ZFU (Zermelo-Fraenkel with {\it Urelemente\/}), and when we restrict the theory to the case which does not consider $m$-atoms, quasi-set theory is essentially equivalent to ZFU, and the corresponding quasi-sets can then be termed `ZFU-sets' (similarly, if also the $M$-atoms are ruled out, the theory collapses into ZFC). The $M$-atoms play the role of the {\it Urelemente\/} in the sense of ZFU.

	In order to preserve the concept of identity for the `well-behaved' objects, an {\it Extensional Equality\/} is introduced for those entities which are not $m$-atoms on the following grounds: for all $x$ and $y$, if they are not $m$-atoms, then $$x =_{E} y := \forall z ( z \in x \leftrightarrow z \in y ) \vee (M(x) \wedge M(y) \wedge x \equiv y)$$

	It is possible to prove that $=_{E}$ has all the properties of classical identity and so these properties hold regarding   $M$-atoms and `sets' (see below). In this paper, all references to `$=$' stand for `$=_E$', and similarly `$\leq$' and `$\geq$' stand, respectively, for `$\leq_E$' and `$\geq_E$'. Among the specific axioms of ${\cal Q}$, few of
them deserve explanation.  The other axioms are adpted from ZFU.

	For instance, to form certain elementary quasi-sets, such as those contaning `two' objects, we cannot use something like the usual `pair axiom', since its standard formulation pressuposes identity; we use the weak relation of indistinguishability instead:

\vspace{0.3cm}
\noindent
[{\em The `Weak-Pair' Axiom\/}] For all $x$ and $y$, there exists a quasi-set whose elements are the indistinguishable objects from either $x$ or $y$. In symbols,\footnote{In all that follows, $\exists_Q$ and $\forall_Q$ are the quantifiers relativized to quasi-sets.} $$\forall x \forall y \exists_{Q} z \forall t (t \in z \leftrightarrow t \equiv x \vee t \equiv y)$$

	Such a quasi-set is denoted by $[x, y]$ and, when $x \equiv y$, we have $[x]$ by definition. We remark that this quasi-set {\it cannot\/} be regarded as the `singleton' of $x$, since its elements are {\it all\/} the objects indistinguishable from $x$, so its `cardinality' (see below) may be greater than $1$. A concept of {\it strong singleton\/}, which plays an important role in the applications of quasi-set theory, may be defined, as we shall mention below.

In ${\cal Q}$ we also assume a Separation Schema, which intuitivelly says that from a quasi-set $x$ and a formula $\alpha(t)$, we obtain a sub-quasi-set of $x$ denoted by $$[t\in x : \alpha(t)].$$

	We use the standard notation with `$\{$' and `$\}$' instead of `$[$' and `$]$' only in the case where the quasi-set is a {\it set\/}.

	It is intuitive that the concept of {\it function\/} cannot also be defined in the standard way, so we introduce a weaker concept of {\it quasi-function\/}, which maps collections of indistinguishable objects into collections of indistinguishable objects; when there are no $m$-atoms involved, the concept is reduced to that of function as usually understood. Relations, however, can be defined in the usual way, although no order relation can be defined on a quasi-set of indistinguishable $m$-atoms, since partial and total orders require antisymmetry, which cannot be stated without identity. Asymmetry also cannot be supposed, for if $x \equiv y$, then for every relation $R$ such that $\langle x, y \rangle \in R$, it follows that $\langle x, y \rangle =_{E} [[x]] =_{E} \langle y, x \rangle \in R$, by force of the axioms of ${\cal Q}$.\footnote{We remark that
$[[x]]$ is the same ($=_{E}$) as $\langle x, x \rangle$ by the Kuratowski's definition.}

	It is possible to define a translation from the language of ZFU into the language of ${\cal Q}$ in such a way that we can obtain a `copy' of ZFU in ${\cal Q}$. In this copy, all the usual mathematical concepts (like those of cardinal, ordinal, etc.) can be defined; the `sets' (in reality, the `${\cal Q}$-sets' which are `copies' of the ZFU-sets) turn out to be those quasi-sets whose transitive closure (this concept is like the usual one) does not contain $m$-atoms.

	Although some authors like Weyl (1949) sustain that (in what regard cardinals and ordinals) ``the concept of ordinal is the primary one'', quantum mechanics seems to present strong arguments for questioning this thesis, and the idea of presenting collections which have a cardinal but not an ordinal is one of the most basic pressupositions of quasi-set theory.

	The concept of {\it quasi-cardinal\/} is taken as primitive in ${\cal Q}$, subject to certain axioms that permit us to operate with quasi-cardinals in a similar way to that of cardinals in standard set theories. Among the axioms for quasi-cardinality, we mention those below, but first we recall that in ${\cal Q}$, $qc(x)$ stands for the `quasi-cardinal' of the quasi-set $x$, while $Z(x)$ says that $x$ is a {\it set\/} (in ${\cal Q}$). Furthermore, $Cd(x)$ and $card(x)$ mean `$x$ is a cardinal' and `the cardinal of $x$' respectively, defined as usual in the
`copy' of ZFU we can define in ${\cal Q}$.

\vspace{0.3cm}
\noindent
[{\it Quasi-cardinality\/}] Every qset has an unique quasi-cardinal which is a cardinal (as defined in the `ZFU-part' of the theory) and, if the quasi-set is in particular a set, then this quasi-cardinal is its cardinal {\em stricto sensu}:\footnote{Then, every quasi-cardinal is a cardinal and
the above expression `there is a unique' makes sense. Furthermore, from the fact that $\emptyset$ is a set, it follows that its quasi-cardinal is 0.} $$\forall_{Q} x \exists_{Q} ! y (Cd(y) \wedge y =_{E} qc(x) \wedge (Z(x) \to y =_{E} card(x)))$$

	${\cal Q}$ still encompasses an axiom which says that if the quasi-cardinal of a quasi-set $x$ is $\alpha$, then for every quasi-cardinal $\beta \leq \alpha$, there is a subquasi-set of $x$ whose quasi-cardinal is $\beta$, where the concept of {\it subquasi-set\/} is like the usual one. In symbols,

\vspace{0.3cm}
\noindent
[{\it The quasi-cardinals of subquasi-sets\/}] $$\forall_{Q} x (qc(x) =_{E} \alpha \to \forall \beta (\beta \leq_{E} \alpha \to \exists_{Q} y (y \subseteq x \wedge qc(y) =_{E} \beta))$$

\vspace{3mm}
Another axiom states that

\vspace{0.3cm}
\noindent
[{\it The quasi-cardinal of the power quasi-set\/}]
$$\forall_{Q} x (qc({\cal P}(x)) =_{E} 2^{qc(x)})$$

\vspace{3mm}
\noindent
where $2^{qc(x)}$ has its usual meaning.

	As remarked above, in ${\cal Q}$ there may exist qsets whose elements are $m$-atoms only, called `pure' qsets. Furthermore, it may be the case that the $m$-atoms of a pure qset $x$ are indistinguishable from one another, in the sense of sharing the indistinguishability relation $\equiv$. In this case, the axiomatics provides the grounds for saying that nothing in the theory can distinguish among the elements of $x$. But, in this case, one could ask what it is that sustains the idea that there is more than one entity in $x$. The answer is obtained through the above mentioned axioms (among others, of course). Since the quasi-cardinal of the power qset of $x$ has quasi-cardinal $2^{qc(x)}$, then if $qc(x) = \alpha$, for every quasi-cardinal $\beta \leq \alpha$ there exists a subquasi-set $y \subseteq x$ such that $qc(y) = \beta$, according to the axiom about the quasi-cardinality of the subquasi-sets. Thus, if $qc(x) = \alpha \not= 0$, the axiomatics does not forbid the existence of $\alpha$  subquasi-sets of $x$ which can be regarded as `singletons'.

	Of course the theory cannot prove that these `unitary' subquasi-sets (supposing now that $qc(x) \geq 2$) are distinct, since we have no way of `identifying' their elements, but qset theory is compatible with this idea.\footnote{The differences among such `unitary' qsets may perhaps be obtained from a distinction between `intensions' and `extensions' of concepts like `electron'. By this way we engage our approach into what Dalla-Chiara and Toraldo di Francia (1993) termed the ``world of intensions''.} In other words, it is consistent with ${\cal Q}$ to maintain that $x$ has $\alpha$ elements, which may be regarded as absolutely indistinguishable objects. Since the elements of $x$ may share the relation
$\equiv$, they may be further understood as belonging to a same `equivalence class' (for instance, being indistinguishable electrons) but in such a way that we cannot assert either that they are identical or that they are distinct from one another (i.e., they act as `identical electrons' in the physicist's jargon).\footnote{The application of this formalism to the
concept of non-individual quantum particles has been proposed in (Krause and French, 1995).}

	We define $x$ and $y$ as {\it similar\/} qsets (in symbols, $Sim(x,y)$) if the elements of one of them are indistinguishable from the elements of the another, that is, $Sim(x,y)$ if and only if $\forall z \forall t (z \in x \wedge t \in y \to z \equiv t)$. Furthermore, $x$ and $y$ are {\it Q-Similar\/} ($QSim(x,y)$) if and only if they are similar and have the same quasi-cardinality. Then, since the quotient qset $x/_{\equiv}$ may be regarded as a collection of equivalence classes of indistinguishable 
objects, then the `weak' axiom of extensionality is:

\vspace{0.3cm}
\noindent
[{\em Weak Extensionality\/}]
\begin{eqnarray}
\forall_{Q} x \forall_{Q} y (\forall z (z \in x/_{\equiv} \to \exists t 
(t \in y/_{\equiv} \wedge \, QSim(z,t)) \wedge \forall t(t \in
y/_{\equiv} \to\nonumber\\
\exists z (z \in  x/_{\equiv} \wedge \, QSim(t,z)))) \to x \equiv y)\nonumber
\end{eqnarray}

	In other words, the axiom says that those qsets that have `the same quantity of elements of the same sort\footnote{In the sense that they belong to the same equivalence class of indistinguishable objects.} are indistinguishable.

	Finally, let us remark that quasi-set theory is equiconsistent with standard set theories (like ZFC) (see (da Costa and Krause, 199*; Krause, 1995; Krause, 199*a)).

\paragraph{2.1 THE QUASI-SET THEORETICAL VERSION OF THE 
INDISTINGUISHABILITY POSTULATE}

\vspace{0.5cm}

\paragraph{ } As we mentioned in the Introduction, the Indistinguishability Postulate intuitively says that permutations of indistinguishable elementary particles cannot be regarded as observable. In order to provide an interpretation of this fact in ${\cal Q}$, let us introduce the following
definition:

\begin{definicao}
A {\em strong singleton} of an object $x$ is a qset $x'$ which satisfies the following property: $$x' \subseteq [x] \wedge qc(x') =_{E} 1.$$
\end{definicao}

	We usually denote the strong singleton $x'$ of $x$ as $St(x)$.

	Thus, according to the above definition, $x'$ is a subquasi-set of $[x]$ (the collection of {\it all\/} objects indistinguishable from $x$) that has just `one element'. This definition makes sense since the quasi-cardinals are cardinals, as entailed by the `Quasicardinality' axiom. Furthermore, it is straightforward to prove in ${\cal Q}$ that for every $x$ there exists a strong singleton of $x$ (Krause, 199*a).

	Then, by recalling further that the qset operations of difference, intersection and union acquire intuitive meanings as in the standard set theories, we can state the following theorem:

\begin{teorema}
{\em [The Indistinguishability Postulate]} 
Let $x$ be a qset and $z$ an $m$-atom such that $z \in x$. If $w \equiv z$,
then $$(x - z') \cup w' \equiv x$$
\end{teorema}

	The proof is an immediate consequence of (mainly) the axiom of Weak Extensionality. By recalling that $z'$ (respect., $w'$) denotes the strong singleton of $z$ (respect., of $w$), then when $w \notin x$ we may interpret the theorem as saying that we have `exchanged' an element of $x$ by an indistinguishable one, and that the resulting fact is that `nothing has occurred at all' (the resulting qsets are indistinguishable). In other words, an indistinguishable from $z$ was exchanged by an indistinguishable from $w$ (provided that $z$ and $w$ are indistinguishable), and nothing occurred with $x$ at all (the remaining quasi-set is indistinguishable fromthe original one).

\paragraph{3. QUASI-SET-THEORETICAL PREDICATE FOR QUANTUM PARTICLES}

\vspace{0.5cm}

\paragraph{ } In this section, we define a quasi-set-theoretical predicate for quantum particle systems.\footnote{This constitutes an extension of Suppes' (1957, 1967) research program which sets as a slogan that to axiomatize a theory is to define a set-theoretical predicate.} In order to deduce, e.g., the quantum statistics, we need to precise the concept of a microobject being in a certain `state'. That is not an easy task, and we shall try to do it as closer to `physicists intuition' as we can.

\begin{definicao}
A structure ${\cal Q}_{bf} = \langle P,{\cal P},F,S,R \rangle$ is a {\em quantum particle system} if and only if the following axioms are satisfied (we shall use $x, y, z$ as variables ranging over $P$ and $p, q, r$ as variables ranging over ${\cal P}$):

\begin{description}
\item [Q1] $P$ is a finite quasi-set.
\item [Q2] $\forall x (x\in P\rightarrow m(x))$. 
\item [Q3] $\forall p (p\in{\cal P}\rightarrow p\subseteq P)$.
\item [Q4] $\forall x (x\in P\wedge F(x) \to m(x) \wedge \forall y 
(y \equiv x \to F(y)))$.
\end{description}

\begin{definicao}
$B(x) =_{\rm def} m(x) \wedge \neg F(x)$\label{bosons}
\end{definicao}

\noindent
{\rm In other words, $F$ is an unary predicate such that $F(x)$ intuitivelly says that $x$ is a `fermion'. If $x$ obeys the predicate $B$, we say that $x$ is a `boson'. Then, definition \ref{bosons} states that any microobject which is not a fermion is a boson, but of course we could modify it conveniently in order to allow also the consideration of other kind of microobjects such as paraparticles (Green, 1953; Hartle and Taylor, 1969).}

\begin{description}
\item [Q5] $S$ is a set endowed with an order relation. The elements of $S$ are called `quantum states'. The elements of $S$ will be denoted by $s, s_{1}, s_{2}, \ldots$. 

\item [Q6] $R$ is a quasi-relation with domain ${\cal P}$ and counter-domain $S$, that is, $R = [[p,s]  : p\in {\cal P}, s\in S]$.

\item [Q7] $\bigcup_{p\in{\cal P}} = P$.

\item [Q8] $\forall p \forall s  
([p,s]\in R \to \forall q (qc(q) > qc(p) \rightarrow [q,s]\not\in R))$.

\item [Q9] $\forall p \forall s \forall q \forall t 
([p,s]\in R\wedge [q,t]\in R\wedge s\neq t\rightarrow p\cap q =
\emptyset)$.

\item [Q10] $\forall x \forall p \forall s (p\in {\cal P} 
\wedge x\in p \wedge s\in S \wedge F(x) \rightarrow ([p,s]\in R 
\rightarrow qc(p)\leq 1))$ (Pauli's Principle).

\end{description}

\end{definicao}

	Axiom {\bf Q1} says that we are considering a finite number of particles only. Axiom {\bf Q2} says that we are concerned only with microobjects, that is, $P$ is a pure quasi-set. Axiom {\bf Q3} says that the elements of ${\cal P}$ are sub-quasi-sets of $P$. Axiom {\bf Q4} says that
every fermion is a microobject and that any object indistinguishable from a fermion is also a fermion. Axiom {\bf Q5} says that we can order quantum states. Axiom {\bf Q6} says that $R$ is a relation whose first elements are sub-quasi-sets of $P$ and whose second elements are quantum states; we can say more about the physical interpretation of $R$. If $[p,s]$ belongs to $R$ and $qc(p) = n$, then we can say that `the quantum state $s$ has occupation number equals to $n$', that is, `there are $n$ quantum particles in the state $s$'. Axiom {\bf Q7} says that every particle of the domain belongs to one of the elements of the choosen collection ${\cal P}$. Axiom {\bf Q8} says that the first elements of the pairs in $R$ have the `maximal number' of elements.  Axiom {\bf Q9} guarantees that one particle cannot be simultaneously in two different quantum states. Axiom {\bf Q10} is our quasi-set theoretical version of Pauli's Exclusion Principle. The relation $R$ provides a way of labelling certain collections of elementary particles. We do not discuss the way of doing this here, since it is outside of our logico-mathematical objectives. One possible way of doing that is by the procedure pointed out in our previous work (Sant'Anna and Krause, 1997). But axiom {\bf Q10} says that the quasi-cardinalities of colections of fermions associated to the same state by $R$ cannot be greater than $1$.

	From our definitions, we have the following results, among others (the proofs are immediate):

\begin{teorema}
Any object indistinguishable from a boson is also a boson.
\end{teorema}

\begin{teorema}
Every microobject is either a fermion or a boson.
\end{teorema}

	In our quasi-set-theoretical framework, the electronic structure of, e.g., the sodium atom ({\it Na\/}) is something like the following. Suppose that $P$, the collection of particles, corresponds to the collection of all electrons in an $Na$-atom. So, it is a pure quasi-set such that $qc(P) = 11$. Let us consider also that the set of states is $S = \{s_1, s_2, s_3,  \ldots, s_{12}\}$. If the quasi-relation $R$ is defined as 

\begin{equation}\label{eq1}
R = [[p_1,s_1],[p_2,s_2],[p_3,s_3],...,[p_{10},s_{10}],[p_{11},s_{11}], 
[p_{12},s_{12}]]
\end{equation}

\noindent
where $p_i\in {\cal P}$ ($i = 1,...,12$) are collections of elements of $P$, and such that $qc(p_1) = ... = q(p_{11}) = 1$, while $qc(p_{12}) = 0$. Thus it results that:

\begin{enumerate}

\item The quasi-sets $p_1,..., p_{11}$ are {\em similar} (cf. section 2);

\item For $i,j = 1...11$, $p_i\equiv p_j$. That is, $p_i$ and $p_j$ are 
{\em indistinguishable} in the sense of the Weak Extensionality Axiom;

\item From the collection of the subsets of $S$, we may define (perhaps by physical means) the set 

\begin{equation}\label{eq2}
E = \{\{s_1,s_2\}, \{s_3,s_4\}, \{s_5,s_6,s_7,s_8,s_9,s_{10}\}, 
\{s_{11},s_{12}\}\}
\end{equation}

\noindent
where each element of $E$, which for the purposes below we shall name $S_{i}$, $i = 1, \ldots, 4$, corresponds to an energy level in the sodium atom. The element $\{s_5,s_6,s_7,s_8,s_9,s_{10}\} \in E$ should correspond to the energy level $2p$, which allows $6$ electrons, once each $s_i$, $i=5,...,10$, is associated to a qset with just $1$ electron (a strong singleton). The element $\{s_{11},s_{12}\}$ coresponds to the energy level $3s$ which allows two electrons but `has' just one electron (recall that $qc(s_{11}) = 1$ and $qc(s_{12}) = 0$). The other two elements of $E$, namely, $\{s_1,s_2\}$ and $\{s_3,s_4\}$, correspond respectively to the 
energy levels $1s$ and $2s$. Then quations (1) and (2) induce the quasi-relation

\begin{eqnarray}
\overline{R} = \left[\left[p_{1} \cup p_{2}, \{s_{1}, s_{2}\}\right], 
\left[p_{3} \cup p_{4}, \{s_{3}, s_{4}\}\right]\right.,\nonumber\\
\left.\left[p_{5} \cup \ldots \cup p_{10}, \{s_{5}, \ldots, s_{10}\}\right],\left[p_{11} \cup p_{12}, \{s_{11}, s_{12}\}\right]\right]
\end{eqnarray}\label{eq3}

\noindent
where $qc(p_{1} \cup p_{2}) =_{E} qc(p_{3} \cup p_{4}) =_{E} 2$, $qc(p_{5} \cup \ldots \cup p_{10}) =_{E} 6$ and $qc(p_{11} \cup p_{12}) =_{E} 1$. The quasi-relation $\overline{R}$ is the quasi-set-theoretical version for the usual rule $1s^2 2s^2 2p^6 3s^1$.
\end{enumerate}

	We observe that this way of seeing the sodium atom is quite `natural' since no identification of the electrons in $P$ is made. The manner to associate each $p_i$ to its `state' in $S$ depends on the so-called `elements of definition'. In other words, we have taken that $R$ is a quasi-relation with counter-domain $S$ (which is a {\it set\/}); mathematically, $S$ aims to provide a way of individualizing the collections. From the `physical' point of view, we may recall that the states are of course not mere collections of particles. As remarked, for instance, by Schr\"odinger (1992), there is a fundamental distinction between `a certain model' (we may say, a certain `quasi-set' $p_{i}$) and `a certain state of this model' (the `pair' $[p_{i}, s_{i}]$). We could interpret a {\it state\/} $s$ as an ordered pair $\langle ex_{s}, in_{s} \rangle$, where $ex_{s}$ is the {\it extension\/} of the concept `state', and $in_{s}$ is its {\it intension\/}; the extension of $s$ may me regarded simply as the collection of objects (elementary particles, in the intended interpretation) which {\it are\/} in that state. The intension of $s$ may then be interpreted as the state's `elements of definition', if we use Schr\"odinger's (1992) terminology, that is, it is the conjunction of the physical properties that characterize the state. The above manner to interpret the quasi-relation $R$ is also strongly connected with the ideas presented in (Dalla-Chiara and Toraldo di Francia, 1993), where it was advanced that `microphysics is a world of intensions'; but we shall leave the details of this point to a future work.

\paragraph{4. QUANTUM STATISTICS}

\vspace{0.5cm}

\paragraph{ } In this section we show how to explain the quantum  distribution functions in our axiomatic framework. In order to see the differences between our approach by using quasi-sets and the 
`classical' one, let us begin by giving an example.

	Suppose that we have two indistinguishable particles labelled 1 and 2 distributed among two distinct states, specified by orthogonal wave-functions $\xi_{\bf 1}$ and $\xi_{\bf 2}$. First of all, let us remark that the use of labels does not compromisse us with any individuation of the particles, for it is precisely whether these labels have meaning in the sense of ascribing individuation to the particles that is questioned here (French, 1989); we intend precisely to show how these labels are dispensable in our formal approach.  Redhead and Teller (1991, 1992) and Teller (1995) sustain that it is precisely this consideration of labels that causes inconvenients in first quantization, since there results `surplus structures' which can be defined in the formalism but that have no counterpart in the physical world. Then they propose to use the second quantized approach, mainly the Fock space formalism, but it was pointed out that even in this case we are not completely free from (at least) conceptual labels (French, Krause and Maidens, 199*).

	But leaving aside these more philosophical questions, let us assume the supposition above, which is closer to the day-to-day manner we consider the quantum mechanical machinery. So, Bose-Einstein statistics permits the following arrangements:

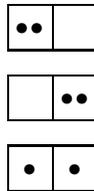
\begin{picture}(100,9)
\put(50,0){\line(1,0){12}}
\put(50,0){\line(0,1){6}}
\put(50,6){\line(1,0){12}}
\put(62,0){\line(0,1){6}}
\put(56,0){\line(0,1){6}}
\put(51,2){$\bullet$}
\put(53,2){$\bullet$}
\end{picture}

\begin{picture}(100,9)
\put(50,0){\line(1,0){12}}
\put(50,0){\line(0,1){6}}
\put(50,6){\line(1,0){12}}
\put(62,0){\line(0,1){6}}
\put(56,0){\line(0,1){6}}
\put(57,2){$\bullet$}
\put(59,2){$\bullet$}
\end{picture}

\begin{picture}(100,9)
\put(50,0){\line(1,0){12}}
\put(50,0){\line(0,1){6}}
\put(50,6){\line(1,0){12}}
\put(62,0){\line(0,1){6}}
\put(56,0){\line(0,1){6}}
\put(52,2){$\bullet$}
\put(58,2){$\bullet$}
\end{picture}

\noindent
whereas Fermi-Dirac statistics allows only

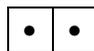
\begin{picture}(100,9)
\put(50,0){\line(1,0){12}}
\put(50,0){\line(0,1){6}}
\put(50,6){\line(1,0){12}}
\put(62,0){\line(0,1){6}}
\put(56,0){\line(0,1){6}}
\put(52,2){$\bullet$}
\put(58,2){$\bullet$}
\end{picture}

	Then, due to the indistinguishability of the particles, the corresponding vectors for the first three cases are {\em chosen} to be respectively

\begin{center}
$\vert \xi^{1}_{\bf 1} \rangle \otimes \vert \xi^{1}_{\bf 2}
\rangle)$

\vspace{2mm}
$\vert \xi^{1}_{\bf 2} \rangle \otimes \vert \xi^{1}_{\bf 1}
\rangle)$

\vspace{2mm}
$\frac{1}{\sqrt{2}} \, (\vert \xi^{1}_{\bf 1} \rangle \otimes
\vert \xi^{2}_{\bf 2}  \rangle + \vert \xi^{1}_{\bf 2} \rangle \otimes \vert \xi^{2}_{\bf 1} \rangle$
\end{center}

\noindent
while for the fourth case, by force of the Pauli's Principle, we choose

\begin{center}
$\frac{1}{\sqrt{2}} \, (\vert \xi^{1}_{\bf 1} \rangle \otimes
\vert \xi^{2}_{\bf 2}  \rangle - \vert \xi^{1}_{\bf 2} \rangle 
\otimes \vert \xi^{2}_{\bf 1} \rangle .$
\end{center}

	We have used the word `choose' above to mark the presence of the Indistinguishability Postulate (IP) here. The labels, initially attached to the particles, are `forgotten' by an adequate selection of states which considers symmetry: any permutation of labels is not regarded as originating a distinct state at all. Even in the anti-symmetrical case this may be asserted, since the expectation value of the state after a permutation is
exactly the same of the state before permutation of the labels, as it is well known.

	In our approach, we deal with collections (quasi-sets) of objects, and not with vectors in an adequate space (which seems more in conformity with the intuitive way of reasoning by means of the boxes above).

	In order to obtain the quantum distribution functions we have considered two assumptions in our axiomatics: First, $S$ is a finite set; that is a necessary hypothesis in the sense that we are interested only in collections of microobjects associated to energy levels of a limited interval of energy. These energy levels may correspond to, in a plausible interpretation, different quantum states. It is quite obvious, therefore, that we are not considering the case of particles under the influence of some kind of non-vanishing potential, like electrons of an atom, for example, since, in this case, different states may correspond to the same energy level. The second assumption is that $P$ is a quasi-set with either only indistinguishable fermions or only indistinguishable bosons.

	By using the terminology of section 3, let us suppose that we have defined a quasi-relation $R$ as equation (1) above, that is, $R = [[p,s] : p\in{\cal P}, s\in S]$. In other words, we are considering a certain collection $P$ of elementary particles subjected to certain states, whose collection we call $S$. Then, for a particular physical situation, we suppose that we are able to select a family  $E = \{S_i\}_{i\in H}$ of subsets of $S$ such that $\bigcap_{i\in H}S_i = \emptyset$, where $H = \{1,2,3, \ldots, n\}$ (which bring us to a situation mathematically similar to equation (2) above). Each $S_{i}$ may be called an {\it energy bin\/} (or {\it macrostate\/}), while each element $s \in S_{i}$ may be called an {\em energy state} (or {\em microstate}). The choice of a particular $S_{i}$ (which is generally done by physical criteria)  particularizes the collections of particles to certain states of interest, namely, the states that belong to the selected set $S_{i}$. To this particular situation $i$, let us suppose that we are considering $qc(P)= \nu_i$ particles, and that $qc(S_i) = k_i$. Then we can define the quasi-relation $R\vert_i = [[p,s] : p\in{\cal P}, s\in S_i]$. 

	Such relation $R\vert_i$ intuitively describes a particular distribution of the $\nu_{i}$ particles into the states in $S_{i}$. So, if we intend  to answer the  fundamental question: `How many ways $I^i$ may we correspond (or distribute) $\nu_i$ indistinguishable bosons (recall that 
$qc(P)\vert_i = \nu_i)$ in the $k_i$ quantum states $s$ of each $S_i$ $(qc(S_i) = k_i)$?' then, since a correspondence between bosons and quantum states is given by the quasi-relations $R\vert_i$, and taking into account that we are talking about bosons, which are not subject to axiom {\bf Q10}, the answer is precisely the quantity of quasi-relations $R\vert_i$ that can be performed.

	Let us explain this result by considering an example. Suppose that 
we have $\nu_i = 5$ indistinguishable bosons to be distributed among $3$
distinct cells (states) $s_1$, $s_2$ and $s_3$. That is, $qc(S_i) = 3$. Since collections of such bosons with the same quasi-cardinality are 
indistinguishable in the sense of the Weak Extensionality Axiom, we shall refer to them by their quasi-cardinalities only. Thus, there are 21  different manners of distributing $5$ indistinguishable bosons into the states $s_1$, $s_2$ and $s_3$, that is, there are 21 {\it possible\/}  quasi-relations $R\vert_{i}^{j}$, $j = 1, \ldots, 21$ which  can be defined
for this particular situation $i$. In the present example, these quasi-relations are shown in the table below, where the integer number entries stand for the quasi-cardinality of the qsets $p_{1}$, $p_{2}$ and $p_{3}$ associated to each $s_i$; the numbers $1, \ldots, 21$ name the relations $R\vert_{i}^{1}$ to $R\vert_{i}^{21}$:

\renewcommand{\arraystretch}{1.5}
\renewcommand{\arraycolsep}{1mm}
\[
\begin{array}{|c|c|c|c|c|c|c|c|c|c|c|c|c|c|c|c|c|c|c|c|c|c|}\hline
\; & ^1 & ^2 & ^3 & ^4 & ^5 & ^6 & ^7 & ^8 & ^9 & ^{10} & ^{11} & ^{12}	& ^{13} & ^{14} &  ^{15} & ^{16} & ^{17} & ^{18} & ^{19} & ^{20} & ^{21}\\ \hline
s_1 & 5 & 4& 4& 3& 3& 3& 2& 2& 2& 2& 1& 1& 1& 1& 1& 0& 0& 0& 0& 0 & 0\\ \hline
s_2 & 0 & 1& 0& 0& 1& 2& 0& 1& 2& 3& 0& 1& 2& 3& 4& 0& 1& 2& 3& 4 & 5\\ \hline
s_3 & 0 & 0& 1& 2& 1& 0& 3& 2& 1& 0& 4& 3& 2& 1& 0& 5& 4& 3& 2& 1 & 0\\ \hline
\end{array}
\]

	For instance, at the column numbered $^3$ we have the quasi-relation
 
$$R\vert_{i}^3 = [[p_{1}, s_{1}], [p_{2}, s_{2}], [p_{3}, s_{3}]]$$

\noindent
where $qc(p_{1}) = 4$, $qc(p_{2}) = 0$ and $qc(p_{3}) = 1$. 

	Then, the number of ways to distribute $\nu_i$ indistinguishable bosons in the $k_i$ quantum states $s$ is 21, a number which is usually obtained by Einstein's equation:

\begin{equation}
I_{bosons}^i = \frac{(k_i+\nu_i-1)!}{(k_i-1)!\nu_i !}.
\end{equation}

	So, Einstein's equation expresses the manner to calculate the number
of possible relations  $R\vert_i$ that can be performed in each situation $i$.\footnote{It is only a simple calculation to prove this result, since the numbers involved in the formula are quasi-cardinals of certain qsets.}
In the same way, if we repeat our calculations for fermions, since they are
subject to axiom {\bf Q10} (there can be no more than one fermion for each quantum state $s\in S$), the answer will be:

\begin{equation}
I_{fermions}^i = \frac{k_i!}{(k_i-\nu_i)!\nu_i !}.
\end{equation}

	An important remark is that usually equations (3) and (4) are obtained by considering that different quantum states correspond to distinct energy levels (Garrod, 1995). Our calculations are more general in the sense that we may have different quantum states at the same energy level as in the case of the sodium atom considered above. In the case of the interpreting quantum states as energy levels, we could consider, for example, that $S$ is an energy interval given by $[0, kT]$, where $k$ is the Boltzmann constant and $T = 300 K$, while each $S_{i}$ corresponds to an energy range of about $10^{-33} J$ (there are $10^{12}$ $S_{i}$'s), and each $S_{i}$ has $10^{19}$ quantum states $s$.

	The total number of microstates corresponding to a given macrostate (energy bin) is then given by:

\begin{equation}
I_{bosons(fermions)} = \prod_i I_{bosons(fermions)}^i
\end{equation}

	Since classical mathematics can be obtained within the scope of quasi-set theory, all the calculations that follow the derivation of the statistics can be performed here as usual. The most probable macrostate will be determined by maximizing $\log I - \alpha N - \beta E$, where $\alpha$ and $\beta$ are Lagrange parameters that have been introduced to take into account the restriction to fix the total particle number $N$, total energy $E$, and $\log$ is the natural logarithm. Thus, it seems clear that we need to define an injective function $f : S \rightarrow {\cal E}$, where ${\cal E}$ is an interval of positive real numbers. Intuitivelly, ${\cal E}$ corresponds to energy.

	In the case of fermions we should maximize the following function $F$:

\begin{eqnarray}
F = \log \left(\prod_i\frac{k_i!}{(k_i-\nu_i)!\nu_i!}\right) -
\alpha\sum_i \nu_i - \beta\sum_i \nu_i\varepsilon_i =\nonumber\\
\sum_i\left[k_i(\log k_i-1)-(k_i-\nu_i)(\log(k_i-\nu_i)-1)\right.\nonumber\\
\left.-\nu_i(\log\nu_i-1)-\alpha\nu_i-\beta\varepsilon_i\nu_i\right],
\end{eqnarray}

\noindent
where $\varepsilon_i$ stands for the energy associated for each $S_i$. It is clear from these calculations that we have used Stirling's approximation, which states that $\log K!\approx K(\log K-1)$ for $K\gg 1$. Nevertheless, such an approximation was used just for bin occupation numbers $\nu_i$ and not for the state ocupation numbers.

	Setting

\begin{equation}
\frac{\partial F}{\partial\nu_i} = 0,
\end{equation}
we get
\begin{equation}
\log[(k_i-\nu_i)/\nu_i] = \alpha+\beta\varepsilon_i,
\end{equation}
which gives
\begin{equation}
\nu_i = \frac{k_i}{e^{\alpha+\beta\varepsilon_i}+1}.\label{nuk}
\end{equation}

	If we assume that the energy differences of the states in the $i$-th bin are negligible, then according to equation (\ref{nuk}), the average occupation of any individual state in that bin is

\begin{equation}
\frac{\nu_i}{k_i} = \frac{1}{e^{\alpha+\beta\varepsilon_i}+1}.
\end{equation}

	Finally, the average ocupation number of the $n$-th single-particle state of energy $\varepsilon_n$ is given by the well-known Fermi-Dirac distribution function:

\begin{equation}
f_{fermions} = \frac{1}{e^{\alpha+\beta\varepsilon_n}+1}.
\end{equation}

	For bosons, the calculations are very similar, and we have the Bose-Einstein distribution function:

\begin{equation}
f_{bosons} = \frac{1}{e^{\alpha+\beta\varepsilon_n}-1}.
\end{equation}

	The physical interpretation of the parameters $\alpha$ and $\beta$ is the usual one. $\beta = 1/kT$, where $k$ is Boltzmann constant and $T$ is the absolute temperature. $\alpha$ is a normalization constant usually refered to as {\em affinity}.

\paragraph{5. THE HELIUM ATOM}

\vspace{0.5cm}

\paragraph{ } The helium atom is probably the simplest realistic situation where the problem of individuality plays an important role. With the identity question put aside, the wave function of the helium atom would be just the product of two hydrogen atom wave functions with $Z = 1$ changed to $Z = 2$. Nevertheless, the space part of the wave function for the case where one of the electrons is in the ground state (100) and the other one is in excited state (nlm) is:

\begin{equation}
\phi({\bf x_1},{\bf x_2}) = \frac{1}{\sqrt{2}}[\psi_{100}({\bf x_1})\psi_{nlm}({\bf x_2})\pm\psi_{100}({\bf x_2})\psi_{nlm}({\bf x_1})],\label{excited}
\end{equation}

\noindent
where the $+$ ($-$) sign is for the spin singlet (triplet)\footnote{Spin singlet refers to total spin zero and spin triplet refers to total spin different of zero.} and ${\bf x_1}$ and ${\bf x_2}$ are the vector positions of both electrons.

	For the ground state, however, the space function must be necessarily symmetric. In this case, the problems regarding identity have no physical effect. The most interesting case is certainly the excited state. Equation (\ref{excited}) reflects our ignorance on which electron is in position ${\bf x_1}$ and which one is in position ${\bf x_2}$. Nevertheless, in the same equation there are terms like $\psi_{100}({\bf x_1})$, which
corresponds to a specific physical property of an individual electron.

	Our quasi-set theoretical interpretation for equation (\ref{excited}) is as follows (it resembles the case of the sodium atom already discussed above). Let $P$ be a pure quasi-set such that $qc(P) = 2$. We intuitivelly interpret the elements of $P$ as electrons of the Helium atom. If $G$ is a unary predicate such that $G(x)$ intuitively says that `$x$ is in the ground state' (the definition of $G$ depends on physical aspects), then, by using the separation axiom of ${\cal Q}$, we obtain the sub-quasi-set $p_1\subseteq P$ defined by

\[p_1 = [x\in P : G(x)].\]

	If we call $p_2 = P-p_1$, then $qc(p_1) = qc(p_2) = 1$. So, the elements of $p$, despite their indistinguishability, are `separated' by their `respective states'. More formally, by calling $g_1$ the ground state and $g_2$ the anther state, we may define

\begin{equation}
R = [[p_1,g_1],[p_2,g_2]].\label{excitedp}
\end{equation}

	It is clear that $g_1$ and $g_2$ may be interpreted respectivelly as $100$ and $nlm$ as above. So, we have arrived to a manner to express the fact that between two objects (the elements of $P$) there is one of them in the ground state, although we cannot identify which one, since the qsets $p_{1}$ and $p_{2}$ are indistinguishable (in the sense of the Weak Extensionality Axiom). In other words, equation (\ref{excitedp}) stands for the situation presented in equation (\ref{excited}). $R$ is a quasi-function, but we still need to show how it evolves in time, that is, we need to explain the sense according to which equation (\ref{excitedp}) plays the role of the wave function given in equation \ref{excited}. Such a topic will be discussed in another paper.

\paragraph{6. FINAL REMARKS AND FURTHER PROBLEMS}

\begin{enumerate}

\item As we mentioned in the Introduction, in a previous work it was presented a manner to cope with collections of `physically' indistinguishable particles in a set-theoretical framework by using hidden variables (Sant'Anna and Krause, 1997). In that paper, such hidden variables were interpreted as `inner properties' which may have no physical interpretation untill now. But, by adapting a kind of realistic point of view, we might say that their meaning is to be achieved by future physics, where perhaps one will be able to distinguish among (at the moment) indistinguishable particles. If we understand that quasi-set theory reflects the idea that individuality is somehow `veiled' for elementary particles, it seems reasonable to make an analogy with the treatment given in terms of hidden variables.

\item According to Hall (1986) it cannot be said that chemistry has been reduced to quantum mechanics since Pauli's exclusion principle cannot be derived from quantum mechanics. Scerri (1995) does not agree with Hall. It is well known that questions regarding reductionism cannot be precisely stated if we do not adopt a set-like-theoretical framework for the physical or chemical theories in question. Reductionism can be defined in terms of isomorphisms, for example (for details see (Suppes, 1967)). On the other hand, we let here as an open problem if it is possible to derive the exclusion principle from an axiomatic framework which is similar to that one presented in this paper. Although we recognise that physicists are generally not concerned with this topic, we agree with Scerri when he says that ``it would be desirable to have a theory which could explain why only anti-symmetric wavefunctions apply to fermions'' [op. cit.].

\item It is usually considered that the interference produced by two light beams is determined by both their mutual coherence and the indistinguishability of the quantum particle paths. The discussion in this paper is focused only on the `corpuscular' features of quantum particles, in the sense that we are not dealing with coherence. Mandel (1991) has proposed a quantitative link between the wave and the particle descriptions by using an adequate decomposition of the density operator. We let such a relation between quasi-sets and coherence as an open problem for future works.

\item It is well known that the unique difference between the electron and the negative muon is their rest mass (Gell-Mann, 1959). That has motivated Dirac (1962) to propose a model for the electron in terms of a membrane so that the muon may be viewed as an electron in an excited `state'. There are two important points about this idea. First, if an electron is a membrane, then it may be individualized in some sense. This point is in agreement with the present paper so as with (Sant'Anna and Krause, 1997) (see item 1 of this section). The second point is a bit more critical. If the muon is an excited electron, it would be possible, by using Schr\"odinger's (1953) terminology, to paint electrons, at least in principle. In other words, it would be possible to mark or individualize electrons by exchanging their intrinsic properties, mainly for those cases where there are just two electrons.

\item As remarked above, this paper is the first one of a series which intends to derive quantum physics into the formalism of quasi-set theory. In this manner, we are walking on the road envised by Yu. Manin (1976) when he proposed that we should search for axioms for collections of indistinguishable objects like elementary particles. Quasi-set theory does this job, and the possibility of obtaining such an alternative way of expressing quantum theory seems to be more in conformity with the possibility that quantum particles are {\em really} indistinguishable objects.

\end{enumerate}

\paragraph{REFERENCES}

\begin{enumerate}

\item Auyang, S.Y., (1995) {\em How is quantum field theory possible?}, New York, Oxford Un. Press.
\item Bitbol, M. and O. Darrigol, (1992), {\em Erwin Sch\"odinger: Philosophy and the birth of quantum mechanics\/}, Paris, Editions Fronti\`ers.
\item Cantor, G., (1955), {\it Contributions to the founding of the theory of transfinite numbers\/}, New York, Dover, [original from 1915].
\item Castellani, E. (ed.), (1997), {\em Interpreting bodies: classical and quantum objects in modern physics\/}, Princeton University Press (1997).
\item da Costa, N.C.A. and D. Krause, (1994), {\em Studia Logica}, {\bf 53}, 533.
\item da Costa, N.C.A. and D. Krause, (199*), `Set-theoretical models for quantum systems', fothcomming in Dalla-Chiara, M.L., R. Giuntini and F. Laudisa (eds.) {\em Philosophy of Science in Florence} (Kluwer Ac. Press). Abstract in the {\em Volume of Abstracts} Xth International Congress of Logic, Methodology and Philosophy of Science, August 19-25, (1995) Florence, pp. 470.
\item Dalla-Chiara, M.L. and G. Toraldo di Francia, (1993), `Individuals, kinds and names in physics', in Corsi, G. et al. (eds.), {\em Bridging the gap: philosophy, mathematics, physics}, (Dordrecht, Kluwer Ac. Press) pp. 261.
\item Dirac, P.A.M., (1962), {\em Proc. Royal Soc. London} {\bf 268 A} 57.
\item Faris, W., (1996), {\it Notices of the AMS\/} {bf 43} 1328.
\item French, S., (1989), {\em Australasian Journal of Philosophy} {\bf 67}, 432.
\item French, S. and M.L.G. Redhead, (1988), {\em British Journal for the Philosophy of Science}, {\bf 39}, 233.
\item French, S., and D. Krause, `The logic of quanta', forthcomming in Cao, T.L. (ed.) {\em Proceedings of the Boston Colloquium for the Philosophy of Science 1996: a historical examination and philosophical reflections on the quantum field theory}, (Cambridge Un. Press).
\item French, S., D. Krause, and A. Maidens, `Quantum vagueness', forthcoming.
\item Garrod, C., (1995), {\it Thermodynamics and statistical
mechanics\/}, Oxford Un. Press.
\item Gell-Mann, M., (1959), {\em Rev. Mod. Phys.} {\bf 31} 384.
\item Green, H.S., (1953), {\em Phys. Rev.} {\bf 90}, 270.
\item Greenberg, O. W. and Messiah, A. M. L., (1964), {\it Phys.
Review\/} {\bf 136 B}, 248.  
\item Hall, P.J., (1986), {\em Synthese} {\bf 69} 273.
\item Hartle, J. and Taylor, J., (1969), {\it Phys. Rev.\/} {\bf 178} 2043. 
\item Krause, D., (1991), {\em J. of Non-Classical Logic\/} {\bf 8}, 9.
\item Krause, D., (1992), {\em Notre Dame Journal of Formal Logic\/} {\bf 33} 402.
\item Krause, D, (1995), `The theories of quasi-sets and ZFC are equiconsistent', in W.A. Carnielli and L.C.P.D. Pereira (eds.) {\em Logic, sets and information\/}, (CLE-UNICAMP) 145.
\item Krause, D. and S. French, (1995), {\em Synthese} {\bf 102} 195.
\item Krause, D. and S. French, (199*), `Opaque predicates and their logic', forthcomming in the {\em Proceedings of the XIth Brazilian Conference on Mathematical Logic}. Abstract in {\em Logic J. of the IGPL} {\bf 5} 3 (1997).
\item Krause, D., (199*a), `Axioms for collections of indistinguishable objects', forthcoming.
\item Krause, D., (199*b), `Remarks on individuation, quantum objects and logic', forthcoming.
\item Lesk, A. M., (1980), {\it J. Phys. A: Math. Gen.\/} {\bf 13} L111. 
\item Mandel, L., (1991), `Coherence and indistinguishability', {\em Optics Letters} {\bf 16} 1882.
\item Manin, Yu. I., (1976), `Problems of present day mathematics: I (Foundations)', in Browder, F.E. (ed.) {\em Proceedings of Symposia in Pure Mathematics} {\bf 28} American Mathematical Society, Providence, 36.
\item Post, H., (1963), {\em The Listener} {\bf 70}, 534. Reprinted in {\em Vedanta for East and West} {\bf 32} (1973) 14.
\item Redhead, M., and P. Teller, (1991), {\em Foundations of Physics\/} {\bf 21}, 43.
\item Redhead, M., and P. Teller, (1992), {\em British Journal for the Philosophy of Science\/} {\bf 43}, 201.
\item Sant'Anna, A.S., and D. Krause, (1997), {\em Found. Phys. Lett.\/} {\bf 10} 409-426.
\item Scerri, E.R., (1995), {\em Synthese} {\bf 102} 165.
\item Schr\"odinger, E., (1952), {\em Science and Humanism\/}, Cambridge University Press.
\item Schr\"odinger, E., (September 1953), {\em Scientific American}, 52.
\item Schr\"odinger, E., (1957), {\em Science Theory and Man\/}, Allen and Unwin.
\item Schr\"odinger, E., (1992), {\em Physique Quantique et R\'epresentation du Monde}, Paris, Seuil.
\item Shoenfield, J.R., (1977), `Axioms of set theory', in Barwise, J., {\em Handbook of mathematical logic\/}, North Holland, 321.
\item Simons, P., (1987), {\em Parts: a Study in Ontology}, Oxford, Clarendom Press.
\item Suppes, P., (1957), {\em Introduction to Logic\/}, Van Nostrand.
\item Suppes, P., (1967), {\em Set-Theoretical Structures in Science\/}, mimeo. Stanford University.
\item Teller, P., (1995), {\em An Interpretive Introduction to Quantum Field Theory\/}, Princeton University Press.
\item Toraldo di Francia, G. (1978), {\em Scientia\/} {\bf 113}, 57.
\item Weyl, H., (1949), {\em Philosophy of Mathematics and Natural
Science\/}, Princeton Un. Press.

\end{enumerate}

\end{document}